\documentclass[english]{article}
\usepackage[T1]{fontenc}
\usepackage[latin9]{inputenc}
\usepackage{float}
\usepackage{amsmath}
\usepackage{amssymb}
\usepackage{graphicx}
\usepackage{esint}

\makeatletter

\providecommand{\tabularnewline}{\\}

\usepackage{amssymb}
\usepackage{latexsym}
\usepackage{epsfig}
\usepackage{float}

\newcommand{\be}{\begin{equation}}
\newcommand{\ee}{\end{equation}}

\makeatother

\usepackage{babel}
\begin{document}
{}~ \hfill\vbox{\hbox{CTP-SCU/2015006}}\break
\vskip 3.0cm
\centerline{\Large \bf Equivalence of open/closed strings}
\vspace*{2.0ex}
\vspace*{10.0ex}
\centerline{\large Peng Wang, Houwen Wu and Haitang Yang}
\vspace*{7.0ex}
\vspace*{4.0ex}
\centerline{\large \it Center for theoretical physics}
\centerline{\large \it Sichuan University}
\centerline{\large \it Chengdu, 610064, China} \vspace*{1.0ex}
\centerline{pengw, hyanga@scu.edu.cn, iverwu@stu.scu.edu.cn}
\vspace*{10.0ex}
\centerline{\bf Abstract} \bigskip \smallskip
In this paper, we prove that the open and closed strings  are $O(D,D)$ equivalent. The equivalence requires an AdS geometry near the boundaries. The $O(D,D)$ invariance is introduced into the Polyakov action by the Tseytlin's action. Traditionally, there exist disconnected open-open or closed-closed configurations in the solution space of the Tseytlin's action. The open-closed configuration is ruled out by the mixed terms of the dual fields. We show that, under some very general guidances, the dual fields are consistently decoupled if and only if the near horizon geometry is $AdS_5$. We then have open-closed and closed-open configurations in different limits of the distances of the $D3$-brane pairs. Inherited from the definition of the theory, these four configurations are of course related to each other by $O(D,D)$ transformations. We therefore conclude that both the open/closed relation and open/closed duality can be derived from $O(D,D)$ symmetries. We then demonstrate the open/closed relation does connect commutative open and closed strings. By analyzing the couplings of the configurations, the low energy effective limits of our results consequently predicts the AdS/CFT correspondence, Higher spin theory, weak gauge/weak gravity duality and a yet to be proposed strong gauge/strong gravity duality. Furthermore, we also have the Seiberg duality and a weak/strong gravitation duality as consequences of  $O(D,D)$ symmetries.

\vfill
\eject
\baselineskip=16pt
\vspace*{10.0ex}
\tableofcontents

\section{Introduction}

It is one of the most important tasks in modern physics to figure
out the relations between gauge theory and gravity, the quantum and
the classical theories. We claim string theory is the TOE, it is therefore
expected to answer these questions. However, the primordial string
theory is only defined in a perturbative way, the Polyakov action.
We thus believe that the answers of these non-perturbative problems
root in M-theory. Before touching the holy grail of M theory, some
non-perturbative facts already emerge from the perturbative theories.
We have the T-duality, which implies the equivalence
of physics between the small and big tori, the S-duality, which unifies
the strong and weak couplings, the U-duality, a combination of T-
and S-dualities. The U-duality is the fundamental symmetry of M theory.
These dualities are known as non-perturbative effects, and exist in
the low energy effective theories. The T-duality can be realized as
the discrete $O\left(d,d;\mathbb{Z}\right)$ symmety. The manifestly
$O\left(d,d;\mathbb{Z}\right)$ invariant action is given in ref.
\cite{Sen:1991cn}. The S-duality manifests $SL\left(2,\mathbb{R}\right)$
symmetry. The corresponding $SL\left(2,\mathbb{R}\right)$ invariant
action can be found in ref. \cite{Sen:1994fa}. Since these dualities
already show up in the low energy effective theories, it is unbelievable
that they cannot be manifested in the world-sheet action. Since we
still do not know how to define M theory, it is highly motivated to
construct intermediate theories between the perturbative string theory
and M theory, by introducing dualities into the world-sheet action.
One can anticipate that these intermediate theories may provide answers
to some of the non-perturbative effects.

Besides these dualities, there exist some amazing relations between
open string and closed string theories. The first one is the open/closed
relation, which connects the open/closed string metrics and couplings.
The low energy effective partner of this relation is the Seiberg-Witten
map \cite{Seiberg:1999vs} between commutative and non-commutative
gauge theories. The second one is the open/closed string duality.
It shows the equivalence between the one-loop open string amplitude
and the tree-level closed string amplitude. Since in low energy limits,
open strings represent gauge theories and closed strings denote gravity,
based on the open/closed string duality, the gauge/gravity duality
is conjectured when $N\rightarrow\infty$ \cite{Horowitz:2006ct}.
The relation between open/closed string duality and gauge/gravity
duality is addressed in \cite{DiVecchia:2003ae}. As a realization
of the gauge/gravity duality, the well-known AdS/CFT \cite{Horowitz:1991cd}
identifies the strong gauge theory and weak gravity. It presents the
correspondence between the quantum gravity in the bulk and gauge theory
on the boundary. There is also a conjectured duality between the strong
gravity and weak gauge theories, the Higher spin theory \cite{Vasiliev:1995dn}.

Let us recall some facts of M theory. M theory has five different
limits or five string theories. We believe that these various string
theories describe the same object from different perspectives. One
of them is the type I string theory, which includes both open and
closed strings. The other four string theories have closed strings
only. Since they describe the same object, there must exist relations
between closed strings and open strings on the level of M theory.
On the other hand, we have known that the U-duality is a fundamental
symmetry of M theory, which is identified as $E_{D}$ group. The U-duality
groups and their maximal subgroups in various dimensions are summarized
in Table (\ref{tab:The-U-duality-group}) \cite{Malek:2012pw}.

\begin{table}[H]
\noindent \begin{centering}
\begin{tabular}{|c|c|c|}
\hline
$D$ & $E_{D}$ & $H_{D}$\tabularnewline
\hline
\hline
3 & $SL\left(3\right)\times SL\left(2\right)$  & $SO\left(3\right)\times SO\left(2\right)$ \tabularnewline
\hline
4 & $SL\left(5\right)$  & $SO\left(5\right)$ \tabularnewline
\hline
5 & $SO\left(5,5\right)$  & $SO\left(5\right)\times SO\left(5\right)$\tabularnewline
\hline
6 & $E_{6}$ & $USp\left(8\right)$\tabularnewline
\hline
7 & $E_{7}$ & $SU\left(8\right)$\tabularnewline
\hline
8 & $E_{8}$ & $SO\left(16\right)$\tabularnewline
\hline
\end{tabular}
\par\end{centering}

\protect\caption{\label{tab:The-U-duality-group}The U-duality group}
\end{table}

It is inspiring to notice that in $D=5$, M theory has $SO\left(5,5\right)$
symmetry. Therefore, we have strong motivation to introduce the $O\left(D,D\right)$
symmetry into the Polyakov action to construct an intermediate theory
between the primordial string theory and M theory. It may capture
some non-perturbative properties of M theory. The $O\left(D,D\right)$
symmetry is a continuous symmetry for non-compact $D$ dimensional
spacetime. After compactifying $d=D-n$ dimensions, the continuous
$O\left(D,D\right)$ breaks into $O\left(n,n\right)\times O\left(d,d;\mathbb{Z}\right)$
group. The discrete $O\left(d,d;\mathbb{Z}\right)$ group is known
as T-duality group in the compactified $d$ dimensional spacetime.
The good news is that there is an available world-sheet action which
manifests $O\left(D,D\right)$ symmetry, the Tseytlin's action. This
action sometimes is also named as double sigma model, built by Tseytlin
\cite{Tseytlin:1990nb} and developed in \cite{Maharana:1992my, Schwarz:1993vs}.

The Tseytlin's action was originally proposed for closed strings, where a set of fields $\tilde X$ dual to the ordinary $X$ is introduced to manifest $O(D,D)$ symmetry.
It was found in \cite{Andriot:2011uh} that non-commutative and commutative
closed string theories can be unified in this theory. The low energy
effective descendant is called double field theory \cite{Siegel:1993xq,Hull:2009mi,Duff:1989tf,Berman:2007xn}.
On the other hand, we showed in \cite{Polyakov:2015wna} that this
action can be perfectly applied on open strings. The non-commutative
and commutative open strings are unified precisely through the open/closed
relation. In the low energy limit, they reduce to the non-commutative
and commutative gauge theories, related by the Seiberg-Witten map
\cite{Seiberg:1999vs}. It is puzzling that, in the open($X$)-open($\tilde X$) and
closed($X$)-closed($\tilde X$) scenarios, there are lots of identical relations. In
these two situations, the dual fields $X$ and $\tilde{X}$ must take
the same boundary conditions, open-open or closed-closed, protected
by the complete form of the boundary conditions. However, it is curious
to ask:
\begin{itemize}
\item Is the open($X$)-closed($\tilde X$) configuration allowed?
\item Why does the open/closed relation connect commutative/non-commutative theories in open-open configuration or closed-closed configuration
but not relate theories in open-closed configuration, just as the name implies?
\end{itemize}
In any case, there are many clues of the equivalence of open strings
and closed strings. The main purpose of this paper is to give 
affirmative answers to these questions. The point is that, there is
a third boundary condition which looks blurry at a first sight, since
it does not fix the string to be open or closed definitely. This actually
is the good news. The bad news is obvious, we can do little without
fixing to open or closed strings. However, once study the complete
form of the boundary conditions carefully, we can find that what forbids
the open-closed configuration is the mixed terms of $X$ and $\tilde{X}$.
It is not hard to understand that if $X$ and $\tilde{X}$ are decoupled
near the boundary, either of them is free to be open or closed. Originally,
the metric in the Tseytlin's action is totally flat, as that in Polyakov
action. But since we are considering an $O(D,D)$ extension of the
Polyakov action, we are certainly allowed to relax the metric to be
spacetime dependent as long as the $O(D,D)$ covariance is preserved.
We show that, the decoupling of $X$ and $\tilde{X}$ happens if and
only if the near boundary geometry is $AdS_{5}$ from some general
considerations! It should be emphasized that the decoupling of the
dual fields only occurs near the boundary. In the bulk, $X$ and $\tilde{X}$
are related by first order differential equations, which is crucial
to have difference to the Polyakov action. We then demonstrate that
under different limits, there are open-closed and closed-open configurations,
all of which are related by $O(D,D)$ transformations. The strength
of the couplings of the low energy effective theories is determined
by the distances between the $D$-brane pairs. We thus can analyze
the symmetries or dualities of the various low energy effective theories.
It turns out that all the relations we mentioned above: AdS/CFT, higher
spin and weak gauge/weak gravity, strong gauge/strong gravity can
be put into this framework and explained by $O(D,D)$ symmetries.  There are also the weak/strong gauge duality known as the Seiberg duality and a weak/strong gravitation duality anticipated from our derivations. Therefore, all the currently known dualities are subsets of the $O(D,D)$ group.

The reminder of this paper is outlined as follows. In section 2, we
show how to realize the open-closed configuration and give the transformations
between all of the four configurations. The consequences of the low
energy effective limits are addressed in section 3. Section 4 is the
summary and discussion.

\section{The equivalence of the open and closed strings}

We start with the Tseytlin's action

\noindent
\begin{equation}
S=-\frac{1}{4\pi\alpha'}\int_{\Sigma}\left(-\partial_{1}X^{M}\mathcal{H}_{MN}\partial_{1}X^{N}+\partial_{1}X^{M}\eta_{MN}\partial_{0}X^{N}\right),\label{eq:Double Sigma Action}
\end{equation}

\noindent where $\partial_{0}=\partial_{\tau}$, $\partial_{1}=\partial_{\sigma}$
and

\noindent
\begin{equation}
\mathcal{H}_{MN}=\left(\begin{array}{cc}
g & -gB^{-1}\\
B^{-1}g & g^{-1}-B^{-1}gB^{-1}
\end{array}\right),\qquad\eta_{MN}=\left(\begin{array}{cc}
0 & 1\\
1 & 0
\end{array}\right),\qquad X^{M}=\left(\begin{array}{c}
X^{i}\\
\tilde{X}_{i}
\end{array}\right),
\end{equation}

\noindent where $M,N=1,2,\ldots,2D$ are $O\left(D,D\right)$ indices,
$g$ is $D$ dimensional spacetime metric and $B$ is the anti-symmetric
Kalb-Ramond field. This action manifests $O\left(D,D\right)$ symmetry,
and the components are invariant under the $O\left(D,D\right)$ rotations:
$\Omega\eta\varOmega^{T}=\eta$. The EOM and boundary conditions can
be obtained by varying the action,

\begin{eqnarray}
\delta S & = & -\frac{1}{2\pi\alpha'}\int_{\Sigma}\delta X^{M}\partial_{1}\left(\mathcal{H}_{MN}\partial_{1}X^{N}-\eta_{MN}\partial_{0}X^{N}\right)\nonumber \\
 &  & +\frac{1}{2\pi\alpha'}\int_{\Sigma}\partial_{1}\left[\delta X^{M}\left(\mathcal{H}_{MN}\partial_{1}X^{N}-\frac{1}{2}\eta_{MN}\partial_{0}X^{N}\right)\right]\nonumber \\
 &  & -\frac{1}{4\pi\alpha'}\int_{\Sigma}\partial_{0}\left[\delta X^{N}\eta_{MN}\partial_{1}X^{M}\right]\nonumber \\
 & & +\frac{1}{4\pi\alpha'}\int_{\Sigma} \delta X^M \partial_1 X^A \partial_M \mathcal{H}_{AN}\partial_1 X^N,\label{eq:variation}
\end{eqnarray}

\noindent where we kept the spacelike boundary for reasons becoming
clear soon. For simplicity, we consider vanishing $B$ field at first.
The EOM is

\begin{equation}
\partial_{1}\left(\mathcal{H}_{MN}\partial_{1}X^{N}-\eta_{MN}\partial_{0}X^{N}\right)=\frac{1}{2}\partial_1 X^A \partial_M \mathcal{H}_{AN}\partial_1 X^N.
\label{eq:old EOM}
\end{equation}

\noindent It turns out that the annoying term on the right hand side of the EOM is irrelevant for almost all of the discussions in this paper. Later we will see that this term  technically narrows down the choices of the metric.  We thus set it vanishing and get

\begin{eqnarray}
g_{ij}\partial_{1}X^{j}-\partial_{0}\tilde{X}_{i} & = & f_{1}\left(\tau\right),\\
g^{ij}\partial_{1}\tilde{X}_{j}-\partial_{0}X^{i} & = & f_{2}\left(\tau\right).
\label{1st EOM}
\end{eqnarray}

\noindent The boundary terms are:

\begin{eqnarray}
\delta X^{i}\left(g_{ij}\partial_{1}X^{j}-\frac{1}{2}\partial_{0}\tilde{X}_{i}\right)+\delta\tilde{X}_{i}\left(g^{ij}\partial_{1}\tilde{X}_{j}-\frac{1}{2}\partial_{0}X^{i}\right)\Bigg|_{\sigma} & = & 0,\label{eq:sigma boundary}\\
\delta X^{i}\partial_{1}\tilde{X}_{i}+\delta\tilde{X}_{i}\partial_{1}X^{i}|_{\tau} & = & 0,\label{eq:tau boundary}
\end{eqnarray}
where $|_{\sigma}$ stands for the timelike boundaries swept by the
end points of the string, and $|_{\tau}$ denotes the initial and
final states. The $|_\tau$ boundary was used to construct D-branes in closed string theory in \cite{Lust:2010iy}.   It is not hard to see that both $X$ and $\tilde{X}$
can represent closed strings (closed-closed). In this scenario, there is no boundary
and we can set $f_{i}(\tau)=0$ by redefining $X$ and $\tilde{X}$.
In \cite{Polyakov:2015wna, DBI}, we showed that $X$ and $\tilde{X}$
can also describe open strings (open-open) by setting

\begin{equation}
\delta X^i|_{\sigma}=\left(g^{ij}\partial_{1}\tilde{X}_{j}-\frac{1}{2}\partial_{0}X^{i}\right)\Bigg|_{\sigma}=0\qquad\Longrightarrow\qquad\partial_{0}X|_{\sigma}=\partial_{1}\tilde{X}|_{\sigma}=0,
\end{equation}
or

\begin{equation}
\delta\tilde{X}_i|_{\sigma}=\left(g_{ij}\partial_{1}X^{j}-\frac{1}{2}\partial_{0}\tilde{X}_{i}\right)\Bigg|_{\sigma}=0\qquad\Longrightarrow\qquad\partial_{1}X|_{\sigma}=\partial_{0}\tilde{X}|_{\sigma}=0.
\end{equation}
In this case, $f_{i}(\tau)$ can also be absorbed into $X$ and $\tilde{X}$. In these two situations, it proves that the Tseytlin's theory can reduce to the Polyakov action.
It is then tempting to ask the question: Is it possible that $X$
represents open strings but $\tilde{X}$ describes closed strings
simultaneously, or vice versa? Since $X$ and $\tilde{X}$ are related
by $O(D,D)$ transformations, if this configuration is permitted,
it becomes possible to explain and furthermore explore the various
dualities and relations between open and closed strings. As we promised
in the introduction, we are going to show that this configuration
does exist.

\subsection{Open-closed configurations}

Referring to eqn. (\ref{eq:sigma boundary}), in addition to the closed-closed
and open-open boundary conditions, there is one more $O(D,D)$ covariant
boundary condition

\begin{equation}
\left(g_{ij}\partial_{1}X^{j}-\frac{1}{2}\partial_{0}\tilde{X}_{i}\right)\Bigg|_{\sigma}=\left(g^{ij}\partial_{1}\tilde{X}_{j}-\frac{1}{2}\partial_{0}X^{i}\right)\Bigg|_{\sigma}=0.
\label{3rd BC}
\end{equation}
This boundary condition is neither open string boundary nor closed
string one. It proves that under this boundary condition, the Tseytlin's action can not reduce to the Polyakov action and it is impossible to remove half of the degrees of freedom by their EOM. This observation implies that \emph{the Tseytlin's theory is more general than the Polyakov one}.  Applying the EOM (\ref{1st EOM}) on the boundary (\ref{3rd BC}), we obtain 
\begin{equation}
f_{1}\left(\tau\right)=-\frac{1}{2} \partial_0 \tilde X\Big|_\sigma,\quad\quad
f_{2}\left(\tau\right)=-\frac{1}{2} \partial_0  X\Big|_\sigma.
\end{equation}
We thus can again absorb $f_{i}\left(\tau\right)$ by shifting $X$ and $\tilde{X}$,

\begin{equation}
\tilde{X}  \to  \tilde{X}-\int d\tau f_{1}\left(\tau\right),\quad\quad
X  \to  X-\int d\tau f_{2}\left(\tau\right).
\end{equation}

\noindent Then the decoupled second order EOM is

\begin{eqnarray}
(\partial_{1}\,^{2}-\partial_{0}\,^{2})X & = & 0,\nonumber \\
(\partial_{1}\,^{2}-\partial_{0}\,^{2})\tilde{X} & = & 0,
\end{eqnarray}

\noindent with the first order constraint,

\begin{eqnarray}
g\partial_{1}X-\partial_{0}\tilde{X} & = & 0,\nonumber \\
g^{-1}\partial_{1}\tilde{X}-\partial_{0}X & = & 0,\label{eq:constraint}
\end{eqnarray}

\noindent and the boundary conditions,

\begin{eqnarray}
\delta X^{i}\left(g_{ij}\partial_{1}X^{j}-\partial_{0}\tilde{X}_{i}\right)+\delta\tilde{X}_{i}\left(g^{ij}\partial_{1}\tilde{X}_{j}-\partial_{0}X^{i}\right)|_{\sigma} & = & 0,\nonumber \\
g_{ij}\delta X^{i}\partial_{0}X^{j}+g^{ij}\delta\tilde{X}_{i}\partial_{0}\tilde{X}_{j}|_{\tau} & = & 0,\label{eq:Boundary Condition}
\end{eqnarray}

\noindent where we applied the constraint (\ref{eq:constraint}) on
the spacelike boundaries. Note due to the shift of $X$ and $\tilde X$, the factor $1/2$ in the boundary conditions (\ref{3rd BC}) disappear, and they are the same as the first order EOM.  To see the picture clearer, we consider the
string propagating between two D-$p$ branes. We use the notations:

\begin{eqnarray}
\mu,\nu,\ldots & = & 0,\ldots,p,\nonumber \\
a,b,\ldots & = & p+1,\ldots,D-1.
\end{eqnarray}

\noindent The boundary conditions (\ref{eq:Boundary Condition}) become

\begin{eqnarray}
\delta X^{a}\left(g_{ab}\partial_{1}X^{b}-\partial_{0}\tilde{X}_{a}\right)+\delta\tilde{X}_{a}\left(g^{ab}\partial_{1}\tilde{X}_{b}-\partial_{0}X^{a}\right)\nonumber \\
+\delta X^{\mu}\left(g_{\mu\nu}\partial_{1}X^{\nu}-\partial_{0}\tilde{X}_{\mu}\right)+\delta\tilde{X}_{\mu}\left(g^{\mu\nu}\partial_{1}\tilde{X}_{\nu}-\partial_{0}X^{\mu}\right)|_{\sigma} & = & 0,\nonumber \\
g_{ab}\delta X^{a}\partial_{0}X^{b}+g^{ab}\delta\tilde{X}_{a}\partial_{0}\tilde{X}_{b}\nonumber \\
+g_{\mu\nu}\delta X^{\mu}\partial_{0}X^{\nu}+g^{\mu\nu}\delta\tilde{X}_{\mu}\partial_{0}\tilde{X}_{\nu}|_{\tau} & = & 0.\label{eq:whole boundary}
\end{eqnarray}

\noindent At a first glance, there can only be closed-closed or open-open
configurations. However, what forbid the open-closed configuration
are the cross terms between $X$ and $\tilde{X}$. Once $X$ and $\tilde{X}$
are decoupled, each of them is free to be open or closed, which leads
to the open-closed configuration! Such a case happens when $g_{\mu\nu}\gg1$
or $g_{\mu\nu}\ll1$, where we used the fact that $g_{ab}$ and $g_{\mu\nu}$
are reciprocal in D-brane theory and $g_{a\mu}=0$. In the original
Tseytlin's theory, the metric was set to be flat. However, we are
trying to generalized the Polyakov action to an $O(D,D)$ invariant
theory. Following the same pattern as Polyakov action to the nonlinear
sigma model, we certainly can go a little bit further to have a nonlinear
double sigma model, as long as the $O(D,D)$ invariance is preserved.
On the other hand, it is well-known that the symmetry group of M-Theory
in five dimensional spacetime is $SO(5,5)$, we thus choose $D=5$.
Moreover, it is generically true that the metric on D-branes is
conformally flat. Therefore, the \emph{near-horizon} metric is almost
fixed to be

\begin{equation}
ds^{2}=f\left(r\right)\eta_{\mu\nu}dx^{\mu}dx^{\nu}+f\left(r\right)^{-1}dr^{2},\qquad g_{\mu\nu}=f\left(r\right)\eta_{\mu\nu},\qquad g_{ab}=f\left(r\right)^{-1},
\end{equation}

\noindent where $dr$ is the normal direction to the D-branes. This
metric is nothing but the geometry of $AdS_{5}$ once requiring it
is consistent with the Einstein equation,

\begin{equation}
ds^{2}=\frac{r^{2}}{c^{2}}\eta_{\mu\nu}dx^{\mu}dx^{\nu}+\frac{c^{2}}{r^{2}}dr^{2},
\end{equation}

\noindent where $c$ is the radius of the AdS. It must be emphasized
that we only need the decoupling of $X$ and $\tilde{X}$ near the
boundary. In the bulk, $X$ and $\tilde{X}$ are coupled. Decoupling
of $X$ and $\tilde{X}$ in the bulk will make the configuration unphysical
as indicated by the EOM (\ref{eq:constraint}). This is very different
from the story of Polyakov action. In Polyakov theory, there does
not exist intermediate mixed open-closed states. Therefore, it is
impossible to have different configurations in different limits. We
now show how the open-closed configuration emerges under the limits.
Substituting the metric into the boundary condition (\ref{eq:whole boundary})
and setting $r\gg c$, we get

\begin{eqnarray}
\delta\tilde{X}_{a}\left(\frac{r^{2}}{c^{2}}\delta^{ab}\partial_{1}\tilde{X}_{b}\right)+\delta X^{\mu}\left(\frac{r^{2}}{c^{2}}\eta_{\mu\nu}\partial_{1}X^{\nu}\right)|_{\sigma} & = & 0,\label{eq:BC r>1 a}\\
\frac{r^{2}}{c^{2}}\delta^{ab}\delta\tilde{X}_{a}\partial_{0}\tilde{X}_{b}+\frac{r^{2}}{c^{2}}\eta_{\mu\nu}\delta X^{\mu}\partial_{0}X^{\nu}|_{\tau} & = & 0.\label{eq:BC r>1 b}
\end{eqnarray}

\noindent From the second term  of eqn. (\ref{eq:BC r>1 a}), we can choose

\begin{equation}
\eta_{\mu\nu}\partial_{1}X^{\nu}|_{\sigma}=0,
\end{equation}

\noindent which makes $X$ is a description of open strings. To cancel the second term on (\ref{eq:BC r>1 b}), we may choose 

\begin{equation}
X\left(\sigma,\tau\right)=X\left(\sigma,\tau+2\pi t\right),
\label{eq:Open periodic}
\end{equation}
which represents a periodic motion of an open string. $t$ runs from $0$ to $\infty$. Bear in mind that
$t$ is not a world-sheet coordinate, but a spacetime coordinate. As a matter of fact, non-compact time evolution also fits the boundary condition, since in this case, as we always do, there is no boundary on the temporal direction for open strings and the second term of eqn. (\ref{eq:BC r>1 b}) identically vanishes. The discussion on this topology is parallel to the compact case. We will mention the differences at the right places. In this paper, we focus on the compact topology in order to make the picture easier to understand.


From the first term of eqn. (\ref{eq:BC r>1 b}), we set

\begin{equation}
\tilde{X}_{b}|_{\tau}=\tilde{Y}_{b},
\end{equation}

\noindent which makes $\tilde{X}$ represent a closed string. The first term of eqn (\ref{eq:BC r>1 a}) vanishes since there is no timelike
boundary for closed strings and we have

\begin{equation}
\tilde{X}(\sigma,\tau)=\tilde{X}(\sigma+2\pi,\tau)
\end{equation}

\noindent The consistent choices for $X^{a}|_\sigma$ and $\tilde{X}_{\mu}|_\tau$
are therefore

\begin{eqnarray}
X^{a}|_{\sigma} & = & Y^{a},\nonumber \\
\eta^{\mu\nu}\partial_{0}\tilde{X}_{\nu}|_{\tau} & = & 0.
\end{eqnarray}

\noindent Therefore, we find the following boundary conditions: for
$X$, we have

\begin{eqnarray}
\textcircled{1}:\qquad\eta_{\mu\nu}\partial_{1}X^{\nu}|_{\sigma} & = & 0,\nonumber \\
X^{a}|_{\sigma} & = & Y^{a},\nonumber \\
X\left(\sigma,\tau\right) & = & X\left(\sigma,\tau+2\pi t\right).
\end{eqnarray}

\begin{figure}[H]
\noindent \begin{centering}
\includegraphics[scale=0.4]{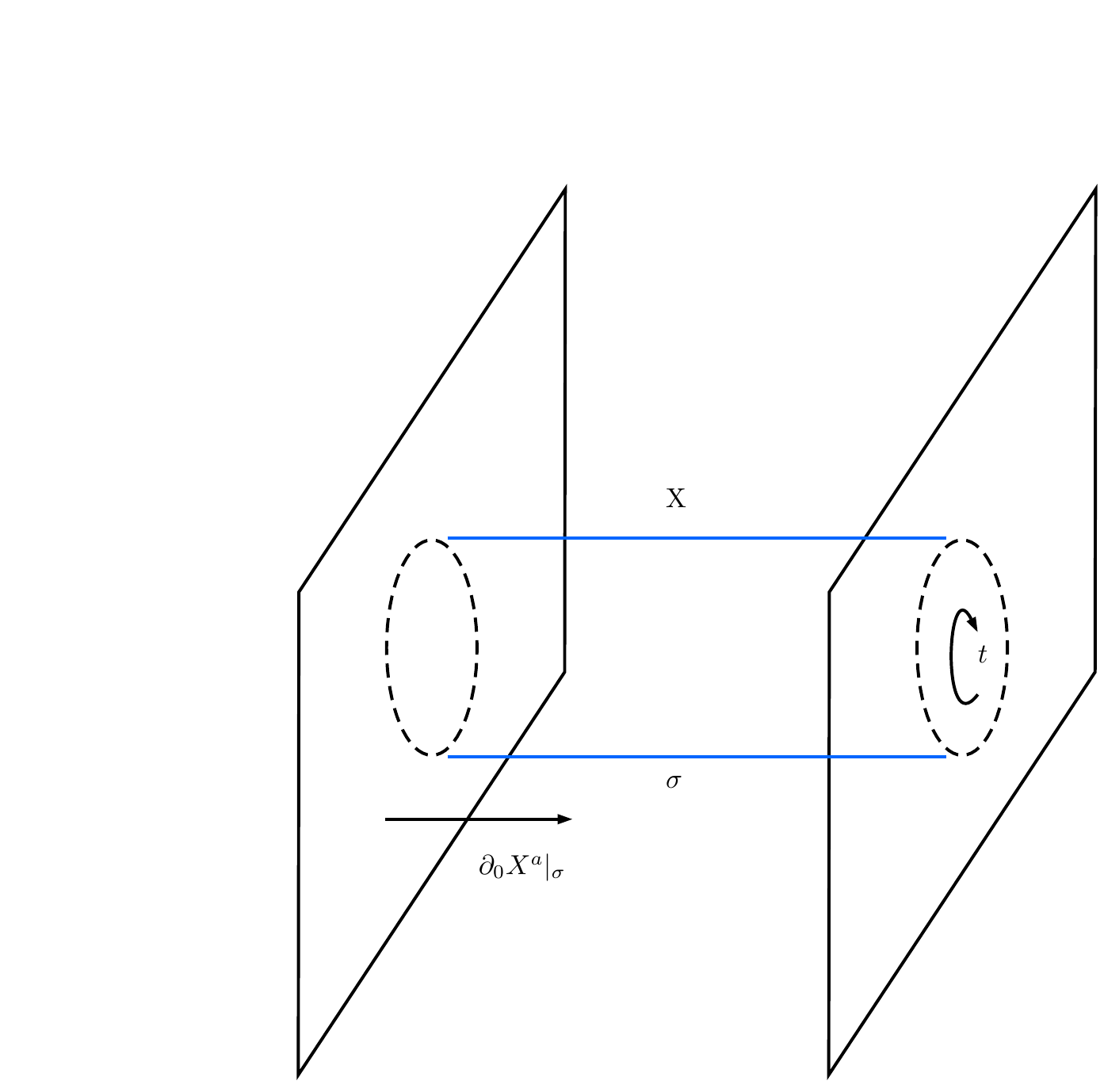}
\par\end{centering}

\protect\caption{\label{fig:Open}$X$ represents a periodic motion of an open string
attached on two D-$3$ branes. The distance of the D-$3$ branes is
the string length. The solid line denotes the string and dashed line
represents the propagation. }
\end{figure}

\noindent The picture respecting these boundary conditions is an open
string ending on two D-$3$ branes and having a periodic motion as depicted
in Fig.\ref{fig:Open}, where the dashed line denotes the $\tau$
direction, the solid line represents the string and the separation
between the two D-$3$ brane is the string length.

\noindent For $\tilde{X}$, we get

\begin{eqnarray}
\textcircled{2}:\qquad\eta^{\mu\nu}\partial_{0}\tilde{X}_{\nu}|_{\tau} & = & 0,\nonumber \\
\tilde{X}_{a}|_{\tau} & = & \tilde{Y}_{a},\nonumber \\
\tilde{X}\left(\sigma,\tau\right) & = & \tilde{X}\left(\sigma+2\pi,\tau\right).
\end{eqnarray}

\begin{figure}[H]
\noindent \begin{centering}
\includegraphics[scale=0.4]{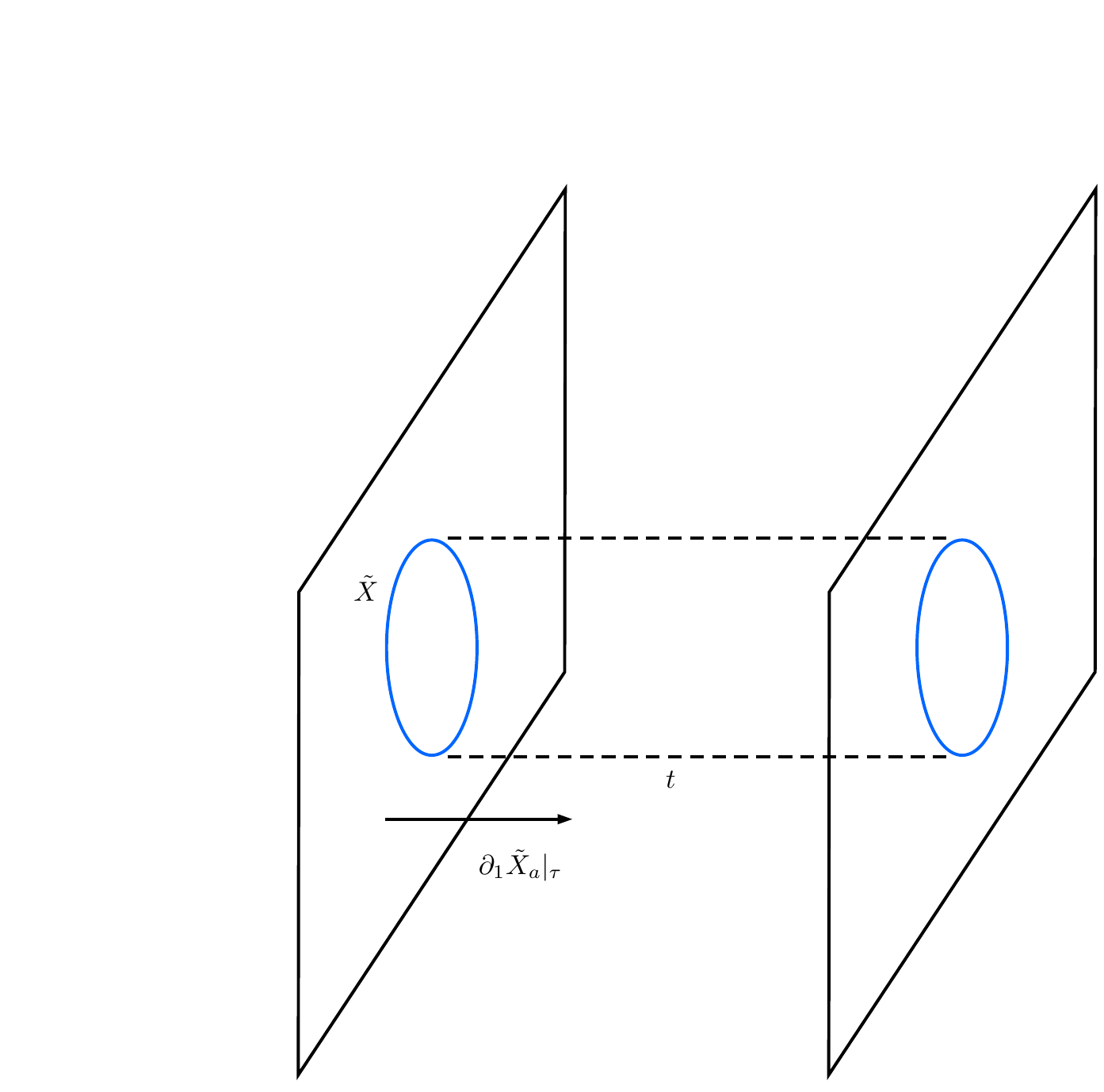}
\par\end{centering}

\protect\caption{\label{fig:Closed}Closed string $\tilde{X}$ propagates from one
D-$3$ brane to another one. The distance of the D-$3$ branes is time
$t$. The solid line denotes the string and dashed line represents
the propagation. }
\end{figure}

\noindent The relevant picture is a closed string propagating from
one D-$3$ brane to another, as shown in Fig. \ref{fig:Closed}. Since
the time of propagation should be the same as $X$, the separation
between the two D-branes is $t$. The equivalence of Fig. \ref{fig:Open}
and Fig. \ref{fig:Closed} was observed long time ago since they have
the same topology. However, it is now clear that this open/closed
duality is a consequence of the $O(D,D)$ symmetry.

Next, we address the opposite limit, $r\ll c$. The metric becomes

\begin{eqnarray}
\delta X^{a}\left(\frac{c^{2}}{r^{2}}\delta_{ab}\partial_{1}X^{b}\right)+\delta\tilde{X}_{\mu}\left(\frac{c^{2}}{r^{2}}\eta^{\mu\nu}\partial_{1}\tilde{X}_{\nu}\right)|_{\sigma} & = & 0,\\
\frac{c^{2}}{r^{2}}\delta_{ab}\delta X^{a}\partial_{0}X^{b}+\frac{c^{2}}{r^{2}}\eta^{\mu\nu}\delta\tilde{X}_{\mu}\partial_{0}\tilde{X}_{\nu}|_{\tau} & = & 0.\label{eq:BC r<1}
\end{eqnarray}

\noindent Since we want to keep the dimensionality of the D-branes unchanged,
under this limit, $X$ is a closed string and $\tilde{X}$ is an open
string. Following the same logic, the boundary conditions as $r\ll c$
are summarized as follows, for $X$,

\begin{eqnarray}
\textcircled{3}:\qquad\eta_{\mu\nu}\partial_{0}X^{\nu}|_{\tau} & = & 0,\nonumber \\
X^{a}|_{\tau} & = & Y^{a},\nonumber \\
X\left(\sigma,\tau\right) & = & X\left(\sigma+2\pi,\tau\right),
\end{eqnarray}

\noindent and for $\tilde{X}$,

\begin{eqnarray}
\textcircled{4}:\qquad\eta^{\mu\nu}\partial_{1}\tilde{X}_{\nu}|_{\sigma} & = & 0,\nonumber \\
\tilde{X}_{a}|_{\sigma} & = & \tilde{Y}_{a},\nonumber \\
\tilde{X}\left(\sigma,\tau\right) & = & \tilde{X}\left(\sigma,\tau+2\pi\frac{1}{t}\right).
\end{eqnarray}

\begin{figure}[H]
\noindent \begin{centering}
\includegraphics[scale=0.6]{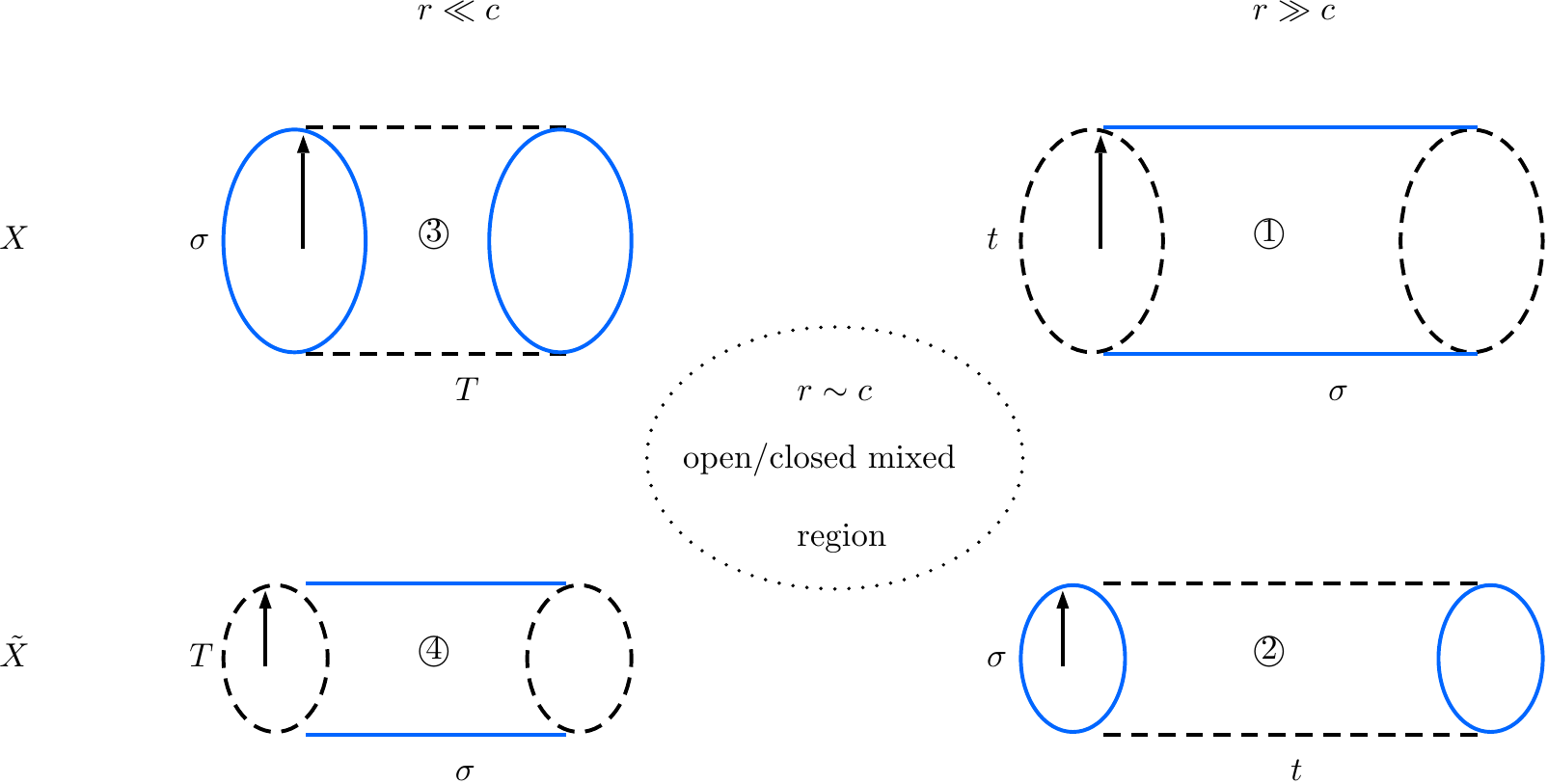}
\par\end{centering}

\protect\caption{\label{fig:4 boundaries} The four configurations in different limits.
The solid lines denote the strings and dashed lines represent the propagations. }
\end{figure}

\noindent Since the limit $r\ll c$ is equivalent to the limit $1/r\gg1/c$,
the distances of the $D$-branes under the two limits are approximate
reciprocals. Thus, in the configuration $\textcircled{4}$, the time
of propagation is $T\backsimeq1/t$, compared with $t$ in the configuration
$\textcircled{1}$. The corresponding pictures of these four configurations
are depicted in Fig. \ref{fig:4 boundaries}.  

It should noted that we identify open or closed strings by the boundary conditions. One is able to do this  near the asymptotic AdS region only. In the $r\sim c$ region, due to the EOM constraint (\ref{eq:constraint}), it is impossible to distinguish open or closed strings. They are entangled and the states are mixed states of open and closed strings. From traditional perspective, the physical picture may look puzzling.

A final remark of this subsection is that, one may notice in eqn. (\ref{eq:Boundary Condition}), on the $|_\tau$ boundary, we applied the first order EOM (\ref{eq:constraint}), which is valid only if the right hand side of EOM (\ref{eq:old EOM}) vanishes, 
\begin{equation}
\partial_1 X^A \partial_M \mathcal{H}_{AN}\partial_1 X^N=0.
\end{equation}
For flat metric, this is  true. However, for generically curved metrics, the story becomes very complicated. Of course, for non-compact topologies, it causes no trouble to our derivations.  For the compact topology, remarkably, for the AdS geometry, since the metric is diagonal and only depends on the radial coordinate, it is not hard to see that only the radial direction $X^a$($\tilde X_a$) is affected. One can easily check that without using the first order EOM (\ref{eq:constraint}), the impacted boundary condition $\delta X^a\partial_1\tilde X_a
+\delta\tilde X_a \partial_1 X^a|_\tau =0$ perfectly fits our identifications of closed strings or open strings under different limits. This observation provides a specific technical reason for the requirement of AdS geometry.

\subsection{Relations of the four configurations}

From the derivations above, we see that the four configurations are
$O(D,D)$ equivalent. To figure out the relations between them, it would be instructive
to put the Kalb-Ramond field into the game. 

After absorbing $f_i(\tau)$, the EOM is

\begin{eqnarray}
g_{ij}\partial_{1}X^{j}-g_{ik}B^{kj}\partial_{1}\tilde{X}_{j}-\partial_{0}\tilde{X}_{i} & = & 0,\nonumber \\
B^{ik}g_{kj}\partial_{1}X^{j}+\left(g^{ij}-B^{ik}g_{kl}B^{lj}\right)\partial_{1}\tilde{X}_{j}-\partial_{0}X^{i} & = & 0.
\end{eqnarray}

\noindent with boundary conditions

\begin{eqnarray}
\delta X^{i}\left(g_{ij}\partial_{1}X^{j}-g_{ik}B^{kj}\partial_{1}\tilde{X}_{j}-\partial_{0}\tilde{X}_{i}\right)+\delta\tilde{X}_{i}
\left(B^{ik}g_{kj}\partial_{1}X^{j}+\left(g^{ij}-B^{ik}g_{kl}B^{lj}\right)\partial_{1}\tilde{X}_{j}-\partial_{0}X^{i}\right)|_{\sigma} & = & 0,\nonumber \\
\delta X^{i}\partial_{1}\tilde{X}_{i}+\delta\tilde{X}_{i}\partial_{1}X^{i}|_{\tau} & = & 0.\nonumber\\
\end{eqnarray}

\noindent It is the right place to introduce the open/closed relation, 

\begin{eqnarray}
g_{ij} & = & \left(\hat{g}-\hat{B}\hat{g}^{-1}\hat{B}\right)_{ij},\nonumber \\
B^{ij} & = & -\left(\frac{1}{\hat{g}+\hat{B}}\hat{B}\frac{1}{\hat{g}-\hat{B}}
\right)^{ij},
\label{eq:open/closed rel}
\end{eqnarray}

\noindent derived from an $O(d,d)$ identification $\eta\mathcal{\hat H}\eta=\mathcal{H}$ as

\begin{equation}
\left(\begin{array}{cc}
\hat{g}-\hat{B}\hat{g}^{-1}\hat{B} & \hat{B}\hat{g}^{-1}\\
-\hat{g}^{-1}\hat{B} & \hat{g}^{-1}
\end{array}\right)=\left(\begin{array}{cc}
g & -gB^{-1}\\
B^{-1}g & g^{-1}-B^{-1}gB^{-1}
\end{array}\right).
\end{equation}

\noindent With these identifications, the EOM is casted into

\begin{eqnarray}
\partial_{1}X^{i} & = & B^{ij}\partial_{1}\tilde{X}_{j}+g^{ij}\partial_{0}\tilde{X}_{j},\nonumber \\
\partial_{1}\tilde{X}_{i} & = & \hat{B}_{ij}\partial_{1}X^{j}+\hat{g}_{ij}\partial_{0}X^{j}.
\end{eqnarray}

\noindent The boundary conditions become

\begin{eqnarray}
\delta X^{i}\left(g_{ij}\partial_{1}X^{j}+\hat{B}_{ik}\hat{g}^{kj}\partial_{1}\tilde{X}_{j}-\partial_{0}\tilde{X}_{i}\right)+\delta\tilde{X}_{i}\left(\hat{g}^{ij}\partial_{1}\tilde{X}_{j}+B^{ik}g_{kj}\partial_{1}X^{j}-\partial_{0}X^{i}\right)|_{\sigma} & = & 0,\nonumber \\
\delta X^{i}\left(\hat{B}_{ij}\partial_{1}X^{j}+\hat{g}_{ij}\partial_{0}X^{j}\right)+\delta\tilde{X}_{i}\left(B^{ij}\partial_{1}\tilde{X}_{j}+g^{ij}\partial_{0}\tilde{X}_{j}\right)|_{\tau} & = & 0,
\end{eqnarray}

\noindent where we used the identity $\hat B\hat g^{-1}=-gB^{-1}$. It is obvious that the system is invariant under\footnote{The actual transformations are $g^{-1}  \leftrightarrow  \hat{g}$ and $B^{-1}  \leftrightarrow  \hat{B}$. But  the confusion is removed by matching the indices.} 

\begin{equation}
X  \leftrightarrow  \tilde{X},\quad
g  \leftrightarrow  \hat{g},\quad
B  \leftrightarrow  \hat{B}.
\end{equation}

\noindent Following the same logic as the case $B=0$, the decoupling of $X$ and $\tilde X$ only occurs in AdS background near the boundaries.  Therefore, we get

\begin{eqnarray}
\delta X^{a}\left(g_{ab}\partial_{1}X^{b}+\hat{B}_{ab}\hat{g}^{bc}\partial_{1}\tilde{X}_{c}-\partial_{0}\tilde{X}_{a}\right)+\delta\tilde{X}_{a}\left(\hat{g}^{ab}\partial_{1}\tilde{X}_{b}+B^{ab}g_{bc}\partial_{1}X^{c}-\partial_{0}X^{a}\right)\nonumber \\
+\delta X^{\mu}\left(g_{\mu\nu}\partial_{1}X^{\nu}+\hat{B}_{\mu\rho}\hat{g}^{\rho\nu}\partial_{1}\tilde{X}_{\nu}-\partial_{0}\tilde{X}_{\mu}\right)+\delta\tilde{X}_{\mu}\left(\hat{g}^{\mu\nu}\partial_{1}\tilde{X}_{\nu}+B^{\mu\rho}g_{\rho\nu}\partial_{1}X^{\nu}-\partial_{0}X^{\mu}\right)|_{\sigma} & = & 0,\nonumber \\
\delta X^{a}\left(\hat{B}_{ab}\partial_{1}X^{b}+\hat{g}_{ab}\partial_{0}X^{b}\right)+\delta\tilde{X}_{a}\left(B^{ab}\partial_{1}\tilde{X}_{b}+g^{ab}\partial_{0}\tilde{X}_{b}\right)\nonumber \\
\delta X^{\mu}\left(\hat{B}_{\mu\nu}\partial_{1}X^{\nu}+\hat{g}_{\mu\nu}\partial_{0}X^{\nu}\right)+\delta\tilde{X}_{\mu}\left(B^{\mu\nu}\partial_{1}\tilde{X}_{\nu}+g^{\mu\nu}\partial_{0}\tilde{X}_{\nu}\right)|_{\tau} & = & 0.
\end{eqnarray}

\noindent Moreover, to have the decoupling in asymptotic regions,  we suppose 
\begin{equation}
B^{-1}g=-\hat{g}^{-1}\hat{B}\sim1.
\end{equation}

\noindent It is clear that $g,B,\hat{g}$ and $\hat B$ have the same monotonicity.
Under the limit $g_{\mu\nu}\gg1$ and $g_{ab}\ll1$, the boundary
conditions are

\begin{eqnarray}
\delta X^{\mu}\left(g_{\mu\nu}\partial_{1}X^{\nu}\right)+\delta\tilde{X}_{a}\left(\hat{g}^{ab}
\partial_{1}\tilde{X}_{b}\right)|_{\sigma} & = & 0,\nonumber \\
\delta X^{\mu}\left(\hat{B}_{\mu\nu}\partial_{1}X^{\nu}+\hat{g}_{\mu\nu}\partial_{0}
X^{\nu}\right)+\delta\tilde{X}_{a}\left(B^{ab}\partial_{1}\tilde{X}_{b}+
g^{ab}\partial_{0}\tilde{X}_{b}\right)|_{\tau} & = & 0.
\end{eqnarray}

\noindent The consistent choices of the boundary conditions are

\begin{eqnarray}
g_{\mu\nu}\partial_{1}X^{\nu}|_{\sigma} & = & 0,\nonumber \\
X^{a}|_{\sigma} & = & Y^{a},\nonumber \\
X\left(\sigma,\tau\right) & = & X\left(\sigma,\tau+2\pi t\right),\nonumber \\
\mathrm{open\; string\; metric}: &  & g.\label{eq:open1}
\end{eqnarray}

\noindent and

\begin{eqnarray}
\hat{g}^{\mu\nu}\partial_{0}\tilde{X}_{\nu}|_{\tau} & = & 0,\nonumber \\
\tilde{X}_{a}|_{\tau} & = & \tilde{Y}_{a},\nonumber \\
\tilde{X}\left(\sigma,\tau\right) & = & \tilde{X}\left(\sigma+2\pi,\tau\right),\nonumber \\
\mathrm{closed\; string\; metric}: &  & \hat{g}.\label{eq:closed1}
\end{eqnarray}

\noindent The boundary condition of $X$ in eqn. (\ref{eq:open1}) precisely represents a commutative open string since  $B$ does not show up and it is not a combination of Neumann and Dirichlet boundary conditions. It is well known that, to have a non-commutative theory, $B$ must be present in the boundary conditions. The metric is nothing but the open string metric $g$. On the other hand, by the similar reason, the boundary condition (\ref{eq:closed1}) of $\tilde X$ describes a commutative closed string. However, it is very instructive to notice that the metric of $\tilde X$ is precisely the closed string metric $\hat g$. Though $B$ or $\hat B$ are not explicitly exhibited in the boundary conditions, they have the same magnitudes of $g$ and $\hat g$. Therefore, the transformation between $g$ and $\hat g$ is still the non-trivial open/closed relation (\ref{eq:open/closed rel}). Thus we have the answer for the second question proposed in the introduction:

\vspace*{3.0ex}

\fbox{\begin{minipage}[t]{1\columnwidth}
The open/closed relation does connect commutative open string  and commutative closed string theories.
\end{minipage}}

\vspace*{3.0ex}

\noindent On the other hand, if we choose $g_{\mu\nu}\ll1$
and $g_{ab}\gg1$, $X$ becomes commutative closed string with metric $g$ and
$\tilde{X}$ becomes commutative open string with metric $\hat g$. 

\begin{figure}[H]
\noindent \begin{centering}
\includegraphics[scale=0.7]{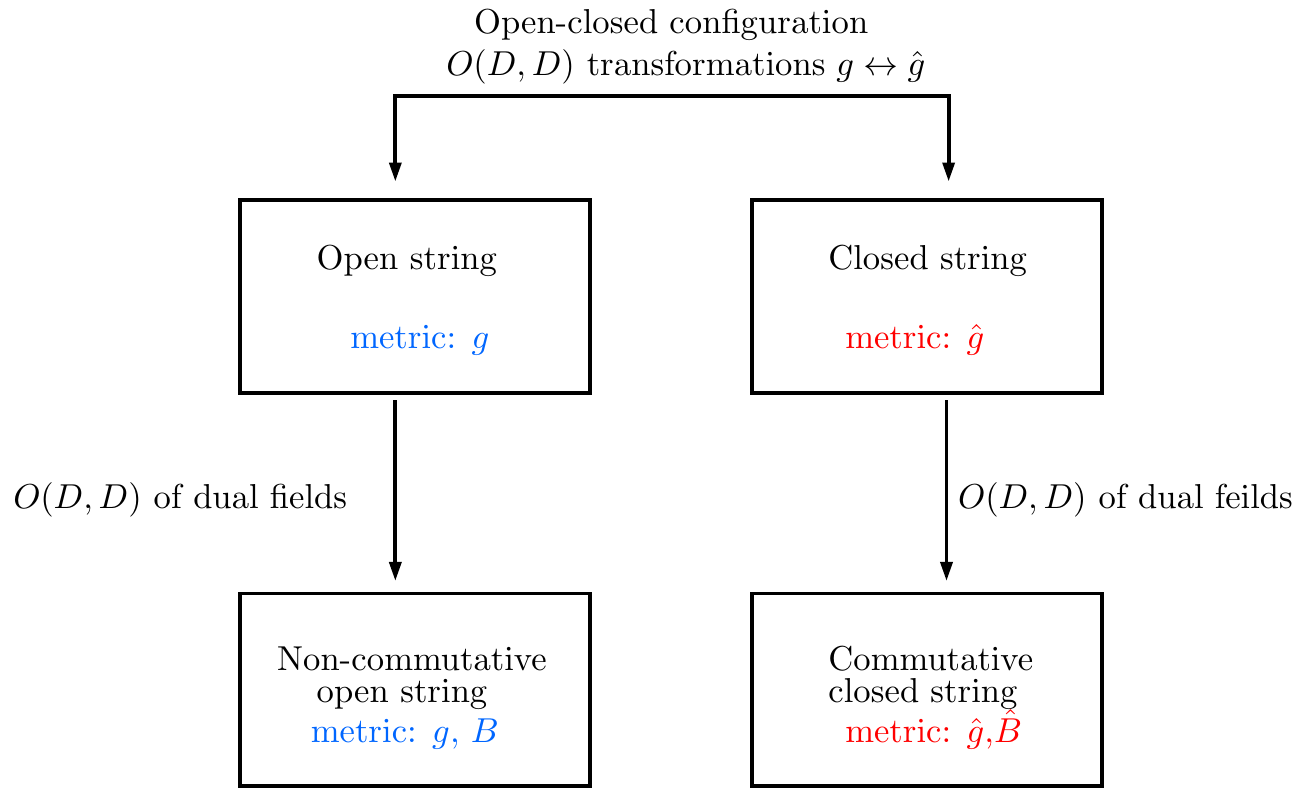}
\par\end{centering}

\protect\caption{\label{fig:com/noncom} Commutative open and closed strings are connected in the asymptotic AdS region by $O(D,D)$.  Either open strings or closed strings rotate back and forth between commutative and non-commutative theories by $O(D,D)$ transformations.}
\end{figure}

Bearing in mind  we have three configurations: open-open, closed-closed and 
open-closed. In the open-open configurations, as we show in \cite{Polyakov:2015wna, DBI}, even with a non-vanishing $B$, there exist both non-commutative and commutative theories for the dual fields. The same story exists in closed-closed configuration. Therefore, we
have the open-closed, commutative/non-commutative relations as in Fig. \ref{fig:com/noncom}.

\begin{figure}[H]
\noindent \begin{centering}
\includegraphics[scale=0.6]{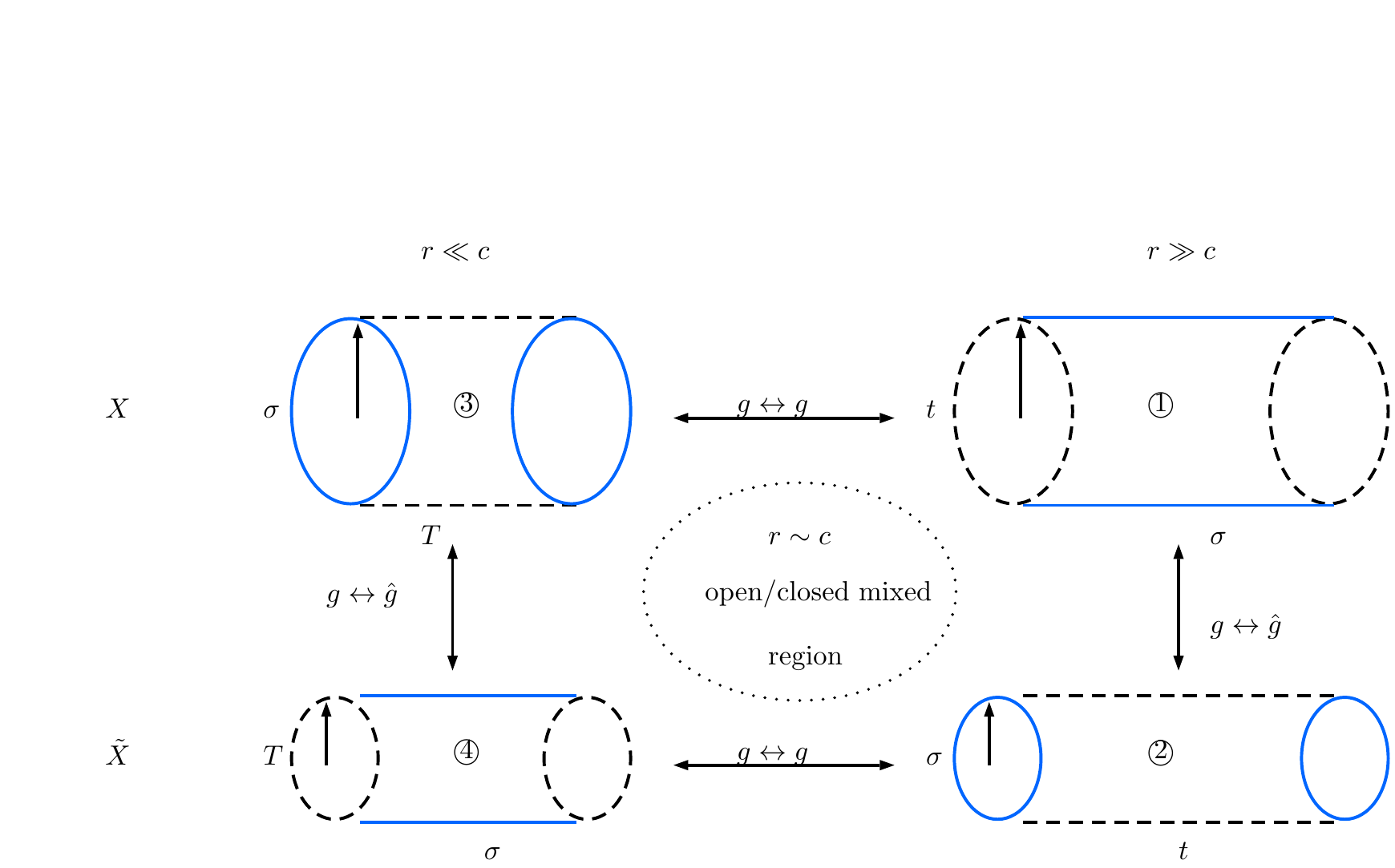}
\par\end{centering}

\protect\caption{\label{fig:four relations} The $O(D,D)$ transformations between
the four states in the open-closed configuration. The solid lines denote the strings and dashed lines represent the propagations. }
\end{figure}

It is now easy to figure out the relations between the four states in the open-closed configuration, as summarized in Fig. \ref{fig:four relations}. 

It worth noting that we only addressed the compact topology of the worldsheet for open strings. We know that the $O(D)$ in $O(D,D)$ is actually non-compact $D(1,D-1)$ group, therefore, when rotating closed strings to compact open string worldsheet, the elements should be in the compact subgroup of $O(D,D)$. On the other hand, as we mentioned before, non-compact open string worldsheet also works perfectly, then the rotations ought to be the non-compact subgroups of $O(D,D)$. It would be of importance to explicitly construct the specific subsets for various manifolds in   future works.  

\section{The low energy effective implications and AdS/CFT }

\begin{figure}[H]
\noindent \begin{centering}
\includegraphics[scale=0.7]{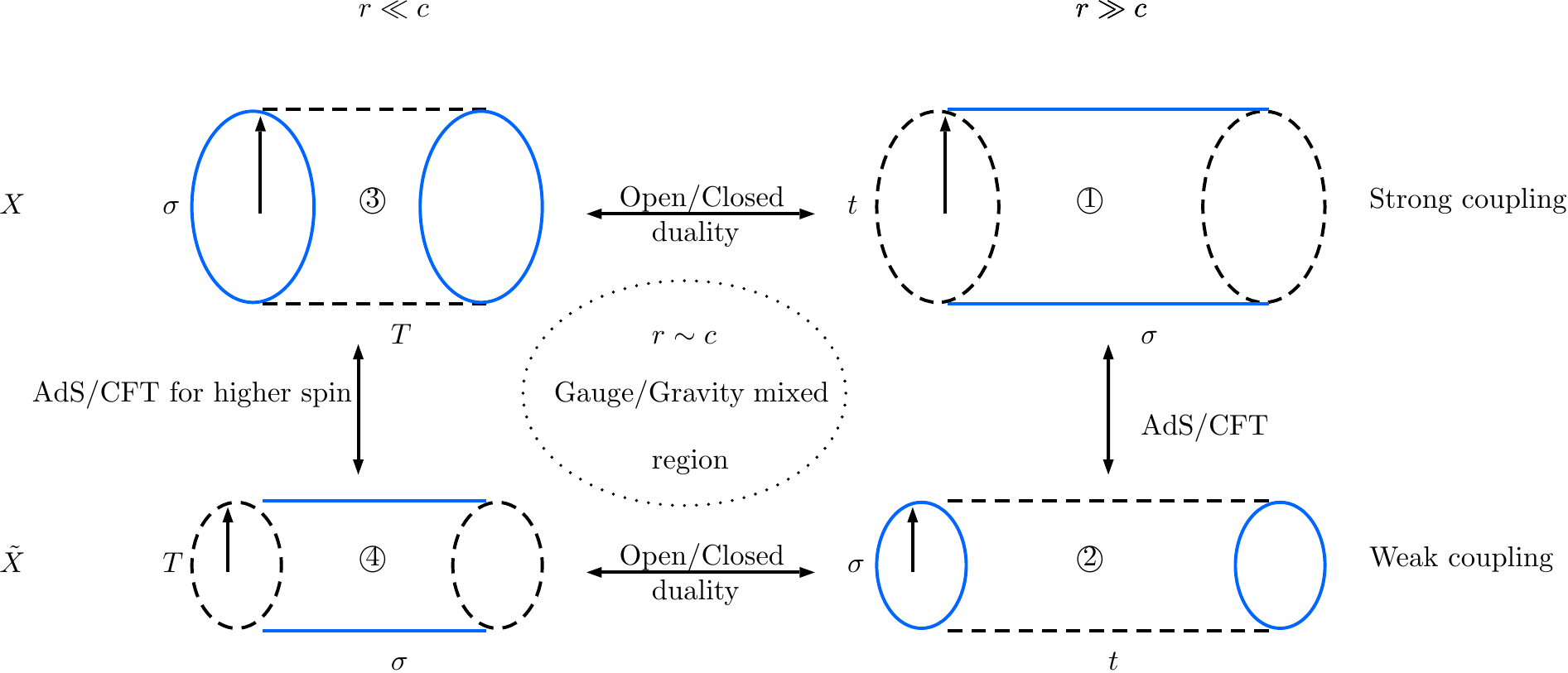}
\par\end{centering}

\protect\caption{\label{fig:low energy relations}Predicted duality web in the low energy limits.}
\end{figure}

In the corresponding low energy limits, the relations among the four
configurations in Fig. \ref{fig:four relations} have important consequences.
For convenience, we list the properties of these configurations in
Table \ref{tab:four string limits}. We know that the couplings of
low energy effective theories are determined by the separation of
$D$-branes. When the separation is large, the propagation of closed
strings becomes far away. The massive modes have no contribution and
we only need to consider the massless sector. But the open string
length is big, the massive modes have significant contributions and
therefore the gauge coupling is large. Oppositely, when the $D$-brane
distance is small, supergravity approximation breaks down, while the
massless sector of open strings dominates. Therefore, the low energy
effective partners of the four configurations are listed in Table
\ref{tab:four low limits} and depicted in Fig. \ref{fig:low energy relations}.
As we explained in the last section, the compactness of the open string worldsheets do not affect the equivalence, our discussions on the low energy effective theories  are general.

In the low energy effective theories, as shown in \cite{Polyakov:2015wna, DBI},
the relation ($\textcircled{1}\leftrightarrow\textcircled{4}$) is related to
the Seiberg-Witten map between the commutative and non-commutative
gauge theories. In string cosmology, the scale-factor duality ($B=0$)
and its extension ($B\not=0$) are realizations of the transformation
between ($\textcircled{2}\leftrightarrow\textcircled{3}$) as addressed
in \cite{Wu:2013sha}, and references therein. Moreover, considering the strength of the couplings, it should represents the S-duality.  The descendant of ($\textcircled{2}\leftrightarrow\textcircled{4}$)
in the low energy limit is conjectured as the weak gauge/weak gravity
duality in literature. To our best knowledge, there is no conjectured
relation for ($\textcircled{1}\leftrightarrow\textcircled{3}$) in
the low energy effective theories, probably because both of them are
strongly coupled and less interesting.

The most interesting duality is the low energy version of ($\textcircled{1}\leftrightarrow\textcircled{2}$)
. From our previous arguments, the $AdS_{5}$ background is a necessity
to decouple $X$ and $\tilde{X}$, which leads to the open-closed
configuration. The decoupling only happens near the boundaries. Their
low energy partners are precisely the weak gravity and strong gauge
theory. All these properties are in agreement with AdS/CFT correspondence!
This observation provides a practical way to prove AdS/CFT correspondence
by getting the relevant low energy effective theories. Moreover, our
results predict the intermediate mixed region. In this region, the
string is neither open nor closed. The dual fields are coupled by
the EOM (\ref{eq:constraint}). It is conceivable that, in the low
energy limit, the gravity and gauge theory are entangled in the intermediate
region as indicated by some recent conjectures. In the AdS/CFT correspondence,
the gauge field concerned is $SU(N)$ Yang-Mills theory. It turns
out that non-abelian and abelian gauge theories are also related by
$O(D,D)$ symmetries \cite{DBI}.

Another duality of the strong gravity and weak gauge theory, deduced
from ($\textcircled{3}\leftrightarrow\textcircled{4}$) represents
the higher spin theory \cite{Vasiliev:1995dn}, attracted tremendous
attention in recent years. Due to the success of AdS/CFT, it is natural
to discuss the dual strongly coupled gravitational theory at short
distance $r/\sqrt{\alpha^{\prime}}\ll1$. Under this limit, the massive
string states become massless since $m^{2}\sim r^{2}/\alpha^{\prime}$,
and string theory reduces to the higher spin gravity. After the reduction,
the modified gauge symmetry involves higher spin fields. Moreover,
it is remarkable that the well-established higher spin theory in AdS$_{3}$/CFT$_{2}$
has a group $SO\left(2,2\right)$ \cite{Engquist:2007yk}, which is
also consistent with our results.

From the off-diagonal configurations, we also observed the existence of strong/weak gauge duality and strong/weak gravity duality. The former one  ($\textcircled{1} \leftrightarrow \textcircled {4}$) is consistent with the electromagnetic duality in abelian theories and its non-abelian extension \cite{Montonen:1977sn, Seiberg:1994pq}. Once all the five dualities exist, one can of course anticipate a duality between the weak and strong gravity, as indicated by ($\textcircled{2} \leftrightarrow \textcircled {3}$).

\begin{table}
\begin{tabular}{|c|c|c|}
\hline
 & $r\ll c$ & $r\gg c$\tabularnewline
\hline
$X$ & $\textcircled{3}$ closed string (in near-horizon geometry) & $\textcircled{1}$ open string (on the boundary)\tabularnewline
\hline
$\tilde{X}$ & $\textcircled{4}$ open string (on the boundary) & $\textcircled{2}$ closed string (in near-horizon geometry)\tabularnewline
\hline
\end{tabular}\protect\caption{\label{tab:four string limits}The properties of the four configurations}
\end{table}

\begin{table}
\begin{tabular}{|c|c|c|}
\hline
 & Short $r$ & Long $r$\tabularnewline
\hline
Open string & $\textcircled{4}$ weak gauge theory & $\textcircled{1}$ strong gauge theory\tabularnewline
\hline
Closed string & $\textcircled{3}$ strong gravitational interaction & $\textcircled{2}$ weak gravitational interaction\tabularnewline
\hline
\end{tabular}\protect\caption{\label{tab:four low limits} The corresponding low energy effective
theories}

\end{table}

\section{Summary and discussion}

In summary, we generalized the Tseytlin\textquoteright s action to
an $O(D,D)$ invariant nonlinear double sigma model. We investigated
the boundary conditions carefully and found, besides the usual open-open
and closed-closed configurations for the dual fields, there does exist
the open-closed configuration. It is remarkable that all the open
and closed configurations are equivalent under $O(D,D)$ symmetries.
This observation proves the equivalence of the open and closed string
descriptions in string theory. To have the open-closed configuration,
it is crucial to decouple the dual fields near the boundary. Surprisingly,
it turns out that the decoupling can only happen in AdS geometry.
From the symmetry group of M theory, we can identify that the AdS
spacetime is precisely $AdS_{5}$. We then explicitly showed that the open/closed relation really connects commutative open strings and commutative closed strings, as its name indicates. We thus proved that the open/closed
relation and open/closed duality are all $O(D,D)$ symmetries. Moreover,
when consider the low energy limits, our results have predictions for
AdS/CFT, higher spin theory, and a weak/weak as well as a strong/strong
dualities between gauge theory and gravity. The Seiberg duality and another weak/strong gravitation duality are also present. The web of the dualities is a realization of $O(D,D)$ symmetry in the low energy effective theories.

There are several remarks we want to address. Though for the open-open and closed-closed configurations, the Tseytlin\textquoteright s action can reduce to the Polyakov action after removing $X$ or $\tilde X$ with the corresponding EOM and boundary conditions, for the open-closed configuration addressed in this paper, effectively, there is no boundary conditions since they can be transformed to the identical form as the EOM. Therefore, one can not remove half of the degrees of freedom and the Tseytlin\textquoteright s action has more physical implications than the Polyakov action. 

The $r\sim c$ region, where the string is in mixed states of open
and closed, is of great importance and interest. It is prompt to study
it carefully and one can expect some non-trivial information can be
extracted.

Constructing the low energy effective theories of the Tseytlin\textquoteright s
action is of course very important. It is not very hard to get the
low energy limits of the open-open and closed-closed configurations.
But the low energy limits of the open-closed configurations are not
straightforward. Moreover, when considering the quantum theory, the
gauge fixing is also subtle. However, once the low energy effective
theories are obtained, we believe all the current dualities can be
understood much better. Especially, since the Gauge/Gravity duality
deals with weakly coupled theories on both sides, one can expect we
may get some instructions to verify the duality.

Our calculation predicts the various dualities only in the background
of AdS. The dS/CFT correspondence proposed in \cite{Strominger:2001pn} is not
compatible with our derivations. However, from the symmetry group
of M theory, it is more precise to restrict our predictions to five dimensional
spacetime. We thus can not exclude the dS/CFT correspondence from
other dimensionality.

It would be of interest to incorporate SUSY into the theory. One may get more information about the required geometry, say, like $AdS_5\times S^5$? It will provide more evidence for the theory.

In this paper, we chose $X$ to be open or closed in different limits to keep the dimensionalities of the D-brane pair. However, at least in pure math, it is perfectly good to choose $X$ to be always open and $\tilde X$ to be always closed in both limits, or vice versa. This choice makes the two D-branes have mutual co-dimensionalities. Does this indicate that there exist some symmetries between different dimensional D-branes?  A related question is that as we know, in the open-open configuration, the dimensionality of the D-brane is determined by the Neumann boundary condition of $X$ or the Dirichlet boundary conditions of $\tilde X$, then which choice is closer to the reality?

The closed string field theory (CSFT) is well-known for its complexity.
The action is non-polynomial, which makes the CSFT is intractable.
On the other hand, the open string field theory (OSFT) is much easier
to deal with, though still very complicated, and has great progresses
in the last twenty years. Our results, the equivalence of the open/closed
string in the asymptotical AdS, provide a possible direction to attack
the CSFT problems by transferring them to the corresponding OSFT problems.

Since we do not know how to define M theory yet, it is a promising
way to generalize the Polyakov theory to $E_{D}$ covariant theories.
It is reasonable to expect some non-perturbative features may be captured
in these extensions. Furthermore, they may be also of help to the construction
of M theory itself.

\vspace{5mm}

\noindent {\bf Acknowledgements}
We are especially indebted to B. Feng for innumerous illuminating discussions and his warm encouragement.  We would like to acknowledge helpful discussions with our collegues Y. He, T. Li, Bo Ning, D. Polyakov and Z. Sun. This work is supported in part by the NSFC (Grant No. 11175039 and 11375121 ) and the Fundamental Research Funds for the Central Universities.  

\end{document}